\documentclass[superscriptaddress,amssymb,aps,twocolumn]{revtex4}

\usepackage{graphicx,dcolumn,bm,amsmath,color,bm}

\newcommand{\eq}[1]{Eq.~(\ref{#1})}
\newcommand{\fig}[1]{Fig.~\ref{#1}}

\newcommand{\eeq}{ \end{equation} }
\newcommand{\beq}{ \begin{equation} }

\newcommand{\eea}{ \end{eqnarray} }
\newcommand{\bea}{ \begin{eqnarray} }

\newcommand{\brr}{ {\bf R} }
\newcommand{\bdr}{ {\bf  r} }

\newcommand{\bz}{ {\bf \hat{z}} }
\newcommand{\bn}{ {\bf \hat{n}} }
\newcommand{\bx}{ {\bf \hat{x}} }
\newcommand{\by}{ {\bf \hat{y}} }

\newcommand{\bal}{ {\bm \hat{\alpha}} }
\newcommand{\bars}{ {\bf \hat{R}} }
\newcommand{\bw}{   \hat{\omega}}
\newcommand{\bwa}{   \hat{\omega}_{1} }
\newcommand{\bwb}{   \hat{\omega}_{2} }
\newcommand{\bwaz}{   \omega_{1z} }
\newcommand{\bwbz}{   \omega_{2z} }
\newcommand{\bwar}{   \omega_{1x} }
\newcommand{\bwbr}{   \omega_{2x} }

\begin{document}

\title{Elastic moduli of  a smectic membrane: a rod-level scaling analysis}
\author{H. H. Wensink and L. Morales Anda} 
\email{wensink@lps.u-psud.fr}
\affiliation{Laboratoire de Physique des Solides - UMR 8502, CNRS \& Universit\'e Paris-Sud, Universit\'{e} Paris-Saclay,  91405 Orsay, France}

\date{\today}

\begin{abstract}

Chiral rodlike colloids exposed to strong depletion attraction may self-assemble into chiral membranes whose twisted director field differs from that of a 3D bulk chiral nematic.  
We formulate a simple microscopic variational theory to determine the elastic moduli of rods assembled into a bidimensional smectic membrane. The approach is based on a simple Onsager-Straley theory for a non-uniform director field that we apply to describe rod twist within the membrane. A microscopic approach enables a detailed estimate of the individual Frank elastic moduli (splay, twist and bend) as well as the twist penetration depth of the smectic membrane in relation to the rod density and shape. We find that the elastic moduli are distinctly different from  those of a bulk nematic fluid, with the splay elasticity being much larger and the curvature elasticity much smaller than for rods assembled in a three-dimensional nematic fluid. We argue that the use of the simplistic one-constant approximation in which all moduli are assumed to be of equal magnitude is not appropriate for modelling the structure-property relation of smectic membranes.  

\end{abstract}

\maketitle

\section{introduction}

Rodlike colloidal particles are capable of assembling into a variety of liquid crystalline mesostructures whose bulk properties depend primarily on the topology of the director field indicating the average direction of rod alignment \cite{dogic-fraden_fil,Ikkala2407}.
In general,  site-specific attractive forces between non-spherical nanoparticles may affect the self-assembly properties  and lead to a wealth of different superstructures \cite{Wang358} such as twisted ribbons of semi-conducting rods \cite{Srivastava1355} or platelets \cite{Janae1701483}.
Mixing rodlike colloids with non-adsorbing polymers or other small depletant particles induces a short-ranged attractive  (`sticky') effective potential between the rods that can be exploited to control the self-assemby morphology \cite{Baranov2010,Sharma2014}. 
 In the case of filamentous {\em fd} virus rods the additional effect of intrinsic particle chirality
strongly impinges  onto the self-assembled mesostructure and drives a vast range of different twisted or chiral structures ranging from smectic membranes to chiral ribbons  \cite{Gibaud2014} and hexagonal nanocrystals displaying screw-dislocations \cite{grelet1}.

Smectic membranes are stabilised by  the strong side-to-side attractions between the rods forcing them to pack into a quasi-bidimensional fluid. An example of such a membrane generated from Monte Carlo computer simulation of hard spherocylinders mixed with penetrable hard spheres acting as a depletant agent  \cite{bolhuis_jcp1997} is shown in \fig{fig1}. In  view of their intrinsic chirality, the rods impart a twist across the membrane which naturally manifests itself at the edge of the membrane given that rod twist is severely penalised near the membrane core where the  curvature is large \cite{kang_sm2016}. The length scale across which the twisting of the rods away from the membrane normal propagates towards the core is referred to the twist penetration depth \cite{gennes-prost}  which has recently be quantified experimentally in smectic membranes composed of {\em fd} rods \cite{barry_jpcb2009}.  Further  important qualitative understanding of the twisted mesostructure and edge fluctuations has been acquired in a number of recent studies based on continuum theories for a smectic membrane \cite{kaplan2010,kaplan2014,kang_sm2016,jia_pelcovits2017}.

The purpose of this paper is to seek a more microscopic rationale for the strongly non-linear twist across a membrane by providing a detailed estimate of the elastic moduli that determine the mesostructure and local alignment of the rods.  Our strategy is to apply Onsager-Straley second-virial theory \cite{straley,wensink_epl2014} originally conceived for bulk nematic fluids with weakly deformed director fields to the case of rods confined to a bidimensional membrane and work out the  relevant distrotion energies. 
Although the second-virial approximation is expected to be more severe for the membrane geometry (even for $L/D \rightarrow \infty$) than for a bulk nematic fluid,  due to the strong quasi-2D confinement the particles experience, it enables the  statistical theory to be rendered analytically tractable  \cite{OdijkLekkerkerker,odijkelastic}. We seek to derive scaling predictions, which should hold for dense membranes in which the local alignment along the concentrically twisting nematic director is asymptotically strong.  The scaling analysis provides us with valuable  information about the way the chiral strength and elastic properties of these membranes depend on the rod concentration, aspect ratio and further details of the interparticle potential. We expect our findings to be valuable for modelling a wide range of fluid membrane structures composed of rodlike mesogens. In particular, our findings allow one to go beyond the simple one-constant approximation in which elastic moduli are assumed equal and their density or temperature-dependence remains elusive. The implications of the derived moduli  on the director twist will be illustrated with some numerical results revealing the main trends upon variation of the three principal length scales that determine the twist angle profile across the membrane, namely the typical distance over which chiral forces propagate (i.e. the inverse pitch), the distance over which twist is expelled to the membrane edge (twist penetration depth) and the size of the membrane itself.

\section{Director deformation in a smectic membrane}
\begin{figure}
\begin{center}
\includegraphics[width= 0.6 \columnwidth]{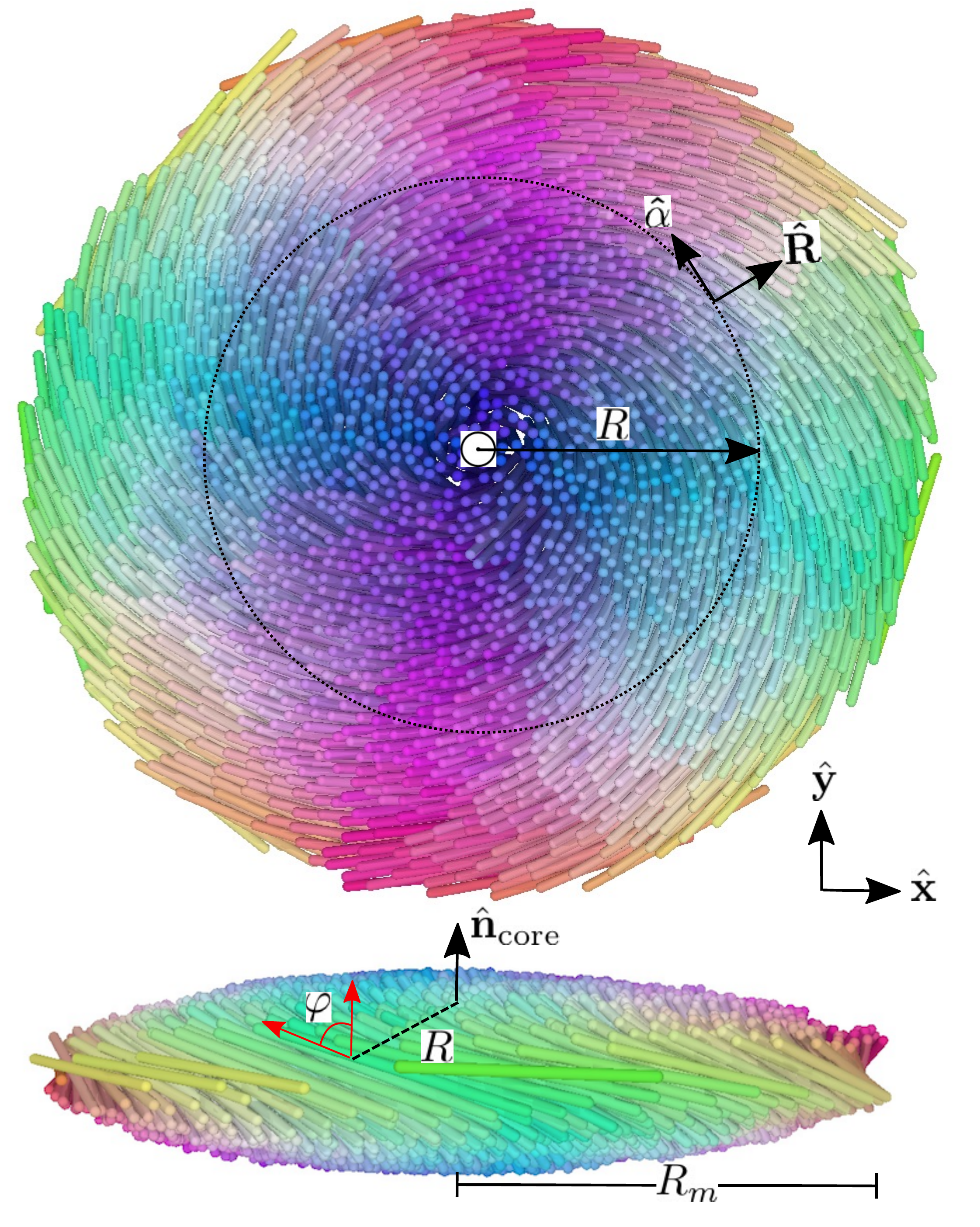}.
\caption{ \label{fig1} Simulation snapshot of a concentrically twisted smectic membrane of radius $R_{m}$ composed of chiral spherocylinders with aspect ratio $\ell = 20$ mixed with non-adsorbing polymers (not shown) providing strong side-to-side depletion attraction between the rods.  The top graph depict the top view, the lower one a side view of the membrane. The director twist, expressed by the twist angle $\varphi$,  is zero at  the membrane  core $\bn _{\rm core} = \bz$ and increases concentrically with radius $R$. Picture courtesy of A. Kuhnhold (Freiburg University, Germany). }
\end{center}
\end{figure}

 Let us consider a membrane assembly in which the rod centers-of-masses are arranged onto a two-dimensional (2D) plane. Assuming the membrane to be circularly symmetric with radius $R_{m}$ we may invoke a cylindrical geometry  with radial coordinate $0<R< R_{m}$ and angle $\alpha$.  Deformations from a uniform director field in which case $\bn = \bz$ are assumed to be concentric and can be described as:
 \beq
 \bn = \cos \psi (R) \cos  \varphi(R) \bz + \cos \psi (R) \sin \varphi(R) \bal + \sin \psi(R) \bars
 \label{tilt}
 \eeq
in terms of a twist angle  $ \varphi $ denoting a twist deflection of the rods with respect to the membrane normal and $\psi$ a splay deformation along the membrane radial vector (see \fig{fig1}).   Let us assume the local rod density within the membrane to be uniform so that the one-body density reads $\rho(R, \bw)  = \rho_{0}f(R,  \bw)$, in terms of the rod number density $\rho_{0}$ denoting the number of rods per area unit, and a three-dimensional rod unit vector $\bw$ distributed along the local director obeying an a priori unknown distribution $f$. 
Without loss of generality, we implicitly fix the thermal energy $k_{B}T$ as our energy unity.  A microscopic expression for the  free energy change per unit area $A$ imparted by a non-uniform director $\bn(\brr)$ (with $\brr$ parameterising 2D space) follows from Straley's extension of the Onsager functional \cite{straley,allenevans} and reads:
\beq
\frac{ F_{el}}{A} = \frac{\rho_{0}^{2}}{4}  \int d \bdr \langle \langle \dot{f}(\bwa) \dot{f}(\bwb) \partial_{\bn}(\bwa) \partial_{\bn}(\bwb)  \Phi(\bdr, \bwa, \bwb)  \rangle \rangle
\eeq 
with $\bdr$ the 2D centre-of-mass distance vector between a rod pair,  $U_{r}$ the associated pair potential  via the Mayer function $\Phi = 1- \exp(- U_{r})$  and $ \partial_{\bn} (\bw_{i}) = (\bdr \cdot \nabla) \bn(\brr) \cdot \bw_{i}  $ a spatial derivative  of the director field. Brackets denote an average over the orientation distribution $f$ via $\langle( \cdot ) \rangle = \int d \bw f(\bw) ( \cdot )$ around the local director and its derivative $\dot{f} = \partial f  / \partial \bw$. If the rods are {\em chiral} then there is a net chiral potential imparting twist across the membrane with associated free energy:
\beq
\frac{ F_{t}}{A} = -\frac{\rho_{0}^{2}}{2} \int d \bdr \langle \langle f(\bwa) \dot{f}(\bwb) \partial_{\bn} (\bwb) \Phi(\bdr, \bwa, \bwb)  \rangle \rangle
\eeq 
For later reference we also formulate the original Onsager excess free energy for a nematic fluid with uniform director field:
\beq
\frac{ F_{ex}}{A} = \frac{\rho_{0}^{2}}{2} \int d \bdr \langle \langle f(\bwa) f(\bwb) \Phi(\bdr, \bwa, \bwb)  \rangle \rangle
\eeq 
Both $F_{ex}$ and $F_{el}$ depend primarily on the repulsive (hard-core) part of the rod interaction potential whereas $F_{t}$ is determined by the microscopic chiral strength of the rods and vanishes if the rods are non-chiral.

It is important to note that the Onsager-Straley theory, by nature of its square-gradient character, represents a local density approximation and is therefore only applicable to situations where gradients in the director are weak on the scale of the typical rod length.  In this context,  we may then assume that the elastic moduli of the membrane are independent of the degree of tilt $\varphi$ and can be estimated from investigating director fluctuations from a uniform (zero tilt) director  field with  $\bn = \bz$.  Linearising \eq{tilt} for small gradients in $\varphi$ and $\psi$ and reverting to a Cartesian frame $\{X\bx,Y\by, Z\bz \}$ enables us to formulate a small twist and splay distortion around the uniform field:  
 \begin{align}
 \bn_{1}  \sim  \bz +   X \left ( \partial_{x} \psi  +  {\mathcal O} (R^{-1}) \right )  \bx \nonumber  \\ 
 \bn_{2}  \sim  \bz +  X \left ( \partial_{x} \varphi + {\mathcal O}(R^{-1} \right  ) \by   
 \label{twangle}
 \end{align}
 The ${\mathcal O} (R^{-1}) $ contributions are curvature corrections that can be identified from  the full cylindrical parameterisation of $\bn$ [\eq{tilt}]. 
 The rod orientation vector can be parameterised as $\bw = \{ \omega_{x}, \omega_{y}, \omega_{z} \} = \{ \sin \theta \sin \phi, \sin \theta \cos \phi, \cos \theta  \} $. Elaborating the free energy expressions above  allows  us to cast the chiral free energy into the following form:
 \beq
 \frac{ F_{t}} {A} \sim  K_{t} (\bn_{2} \cdot \nabla \times \bn_{2}) \sim K_{t} \left ( \partial_{R} \varphi + (2R)^{-1} \sin 2 \varphi   \right ) 
 \label{ft}
 \eeq
Likewise we write for the elastic free energy containing a splay and a twist contribution:
\begin{align}
\frac{ F_{el}}{A} & \sim  \frac{K_{1}}{2}  (\nabla \cdot \bn_{1})^{2}  + \frac{K_{2}}{2} (\bn_{2} \cdot \nabla \times \bn_{2})^{2} \nonumber \\
& \sim  \frac{ K_{1}}{2} (\partial_{R} \psi + R^{-1} \psi)^{2} + \frac{K_{2}}{2} (\partial_{R} \varphi + (2R)^{-1} \sin 2\varphi   )^{2}  
 \end{align}
where the curvature terms have been obtained from replacing $\bn_{1}$ and $\bn_{2}$  by \eq{tilt} assuming weak splay $\psi \ll 1$, so that $ \bn \sim  \cos  \varphi \bz +  \sin \varphi \bal + \psi \bars$. 
The reference excess free energy for the untwisted fluid membrane simply reads $ F_{ex}/A \sim K_{0}$.
The moduli $K_{n}$ are associated with the propensity of the rods to align ($K_{0}$), transmit chiral torques ($K_{t}$) and provide elastic resistance to fluctuations of the director field  ($K_{1}$ and  $K_{2}$). They are defined in terms of the following orientational averages: 
 \begin{align}
 K_{0}  & = -\frac{\rho_{0}^{2}}{2} \langle \langle f( \bwa)f( \bwb)  M_{0} \rangle \rangle \nonumber \\
 K_{t} & =  -\frac{\rho_{0}^{2}}{2} \langle \langle f( \bwa) \dot{f}( \bwb) \omega_{2y} M_{1}  \rangle \rangle \nonumber \\ 
 K_{1} &=  -\frac{\rho_{0}^{2}}{2} \langle \langle \dot{f}( \bwa) \dot{f}( \bwb) \omega_{1x} \omega_{2x} M_{2} \rangle \rangle \nonumber \\
 K_{2} &=  -\frac{\rho_{0}^{2}}{2} \langle \langle \dot{f}( \bwa) \dot{f}( \bwb)  \omega_{1y} \omega_{2y} M_{2} \rangle \rangle 
 \label{moddef}
 \end{align}
The kernels $M_{n }$  are defined in terms of weighted second-virial integrals thus giving the moduli a 
distinct microscopic basis:
\beq
M_{n}  =  \int d \bdr  ({\bf r} \cdot x) ^{n}   \Phi(\bdr, \bwa, \bwb)
\label{kerm}
\eeq
In the case of {\em hard} rods, where the pair potential is infinite if the cores overlap and zero if they do not,  these expressions can be identified as generalised excluded {\em areas}. The overlap integrals \eq{kerm}  can be deduced analytically. Considering hard cylinders and retaining the leading order expressions in the limit of infinite aspect ratio $L/D \rightarrow \infty$  we arrive at the following expressions:
\begin{align}
 M_{0} & \sim LD W^{-} |\sin \gamma |  \nonumber \\
 M_{2}  & \sim \frac{L^{3}D}{12} |\sin \gamma | \frac{(\bwaz \bwbr- \bwbz \bwar )^{2}}{\bwaz^{2} \bwbz^{2} } \nonumber \\ 
& \times  \left [ (\bwaz^{2} + \bwbz^{2}) W^{-}  +  \bwaz \bwbz W^{+} \right ]
\label{meven}
\end{align}
with $ W^{\pm} = ( |\bwaz - \bwbz | \pm | \bwaz+ \bwbz | ) / \bwaz \bwbz  $. 
The first contribution is identical to the one derived in Ref. \cite{shundyakinterface} where it was used to determine the interfacial structure of a rod fluid, whereas the second expression is needed for the calculation of the elastic moduli.

\subsection{Bend elasticity}
\noindent 
In addition to the twist and splay elastcity,  there is a bend elastic contribution which is due {\em solely} to the local curvature of the twisted membrane if the  tilt angle $\varphi$ is nonzero.  A  perfectly aligned smectic membrane does not experience any bend elasticity. The bend term is proportional to the square of the tilt angle $\varphi$ leading to a curvature correction to  \eq{twangle}. A simple geometric analysis of the curvature-induced fluctuation of the local director gives: 
 \beq
 \bn_{3} \sim  \bz +    \frac{Y}{2} \frac{(\sin \varphi)^{2} }{R}  \bx
 \eeq
 where $\varphi \ll 1$.  This translates into a bend elastic free energy 
 \beq
\frac{  F_{el}}{A}\sim \frac{K_{3}}{2} (\bn_{3} \times \nabla \times \bn_{3})^{2} \sim \frac{ K_{3}}{2}  \frac{ (\sin \varphi )^{4} }{ R^{2}} 
\label{k3}
\eeq q
with the corresponding modulus given by:
\beq
 K_{3} =  -\frac{\rho_{0}^{2}}{8} \langle \langle \dot{f}( \bwa) \dot{f}( \bwb) \omega_{1y} \omega_{2y} M_{2} \rangle \rangle = \frac{1}{4} K_{2}
\eeq
Clearly, the bend elasticity strongly expels twist near the core of the membrane where the curvature $1/R$ is large.

\subsection{Chiral twisting strength }

 \noindent 
The torque-field contribution $M_{1}$ can be estimated by considering a weakly chiral pair potential $U_{c}$ described by some arbitrary but short-ranged spatial decay function $g(r)$ describing the range over which chiral forces interact and chiral amplitude $\varepsilon_{c} $ much smaller than the thermal energy. This potential takes the following generic form \cite{goossens1971}:
\beq
U_{c} \sim  \varepsilon_{c} g( r)(\bwa \times \bwb \cdot   \bx) (\bwa \cdot \bwb)
\label{uchi}
\eeq
Adopting a simple van der Waals approach allows us to approximate $\Phi \sim - U_{c}$ so that the strength of the microscopic torques transmitted by the rods  is quantified by:
\begin{align}
M_{1}  \sim    \bar{\varepsilon}_{c} ([ ({\bf I} - \bz \otimes \bz) \cdot (\bwa \times \bwb) ] \cdot \bx ) (\bwa \cdot \bwb)
\end{align}
where the tensor  (with ${\bf I}$ the unity matrix)  imparts the projection of a 3D vector (in this case the cross product of the rod vectors) onto the 2D lab frame of the membrane. The integrated chiral amplitude  $\bar{\varepsilon}_{c}$ is distinctly different from that of a  3D cholesteric system as it implicitly encodes the geometric confinement since $\bdr$ is a 2D vector:
\beq
 \bar{\varepsilon}_{c}  =  \varepsilon_{c} \int d \bdr ( \bdr \cdot \bx)   g( r)
 \eeq
The chiral potential drives the twisting of the membrane and $K_{t}$  provides an explicit link between the coarse-grained effective torque-field $F_{t}$ and the range and amplitude  of the chiral pair potential.

\subsection{Gaussian approximation}

Proceeding toward an explicit calculation of the moduli $K_{n}$ we consider a simple Gaussian Ansatz for the orientational fluctuations around the local director \cite{Vroege92}:
 \beq
 f_{G}( \bw) \sim \frac{\sigma}{4 \pi}  e^{ - \frac{1}{2} \sigma \theta^{2}}
 \eeq
in terms of a variational order parameter  $\sigma  \gg 1$ describing weak  fluctuations  in the polar angle $\theta = \cos ^{-1} (\bw \cdot\bn) \ll 1$ of each rod. Its mirror form $\theta \rightarrow \pi - \theta$ applies to rods pointing anti-parallel to the local director (assuming both have equal probability, i.e. there is no polarity). The derivative featuring in the expressions for the elastic moduli \eq{moddef} simply reads $\dot{f}_{G} = \sigma f_{G}$.  In the weak fluctuation limit  $\theta \ll 1$ we write up to leading order  $\bw = \{  \theta \sin \phi , \theta  \cos\phi , 1 \}$. If we further assume zero local biaxiality across the membrane then $f$ does not depend on the azimuthal angle $\phi$. We can mitigate the azimuthal dependency by writing $\phi_{2}= \phi_{1} + \Delta \phi $ and pre-integrating the kernels  via $(2 \pi)^{-1} \int_{0}^{2 \pi} d \phi_{1}$.
Some basic trigonometric manipulations allow us to express the kernels in terms of the relevant angular variables (retaining the leading order terms only). Tedious but straightforward algebra then gives us the following asymptotic expressions for the angular dependencies:
\begin{align}
  M_{0}    & \sim  -2 LD |\gamma| \nonumber \\
  \omega_{2y} M_{1}   & \sim  - \frac{1}{2}  \bar{\varepsilon}_{c}  \theta_{2}^{2}  \nonumber \\ 
  \omega_{1y} \omega_{2y} M_{2}   &  \sim   \frac{L^{3}D}{24} \left (  |\gamma | \theta_{1}^{2} \theta_{2}^{2} \cos 2 \Delta \phi - |\gamma |  \theta_{1}^{3} \theta_{2} \cos \Delta \phi \right )    \nonumber \\ 
    \bwar \bwbr M_{2}  &  \sim   \frac{L^{3}D}{12} \nonumber \\ 
 & \times \left [  |\gamma | \theta_{1}^{2} \theta_{2}^{2} \left (1 + \frac{1}{2} \cos 2 \Delta \phi \right )   -\frac{3}{2} | \gamma | \theta_{1}^{3}\theta_{2} \cos \Delta \phi   \right ]
  \end{align}
Throughout the analysis we fix the rod diameter $D$ as unit length in terms of a dimensionless density $\rho_{0} \rightarrow \rho_{0} D^{2}$, rod length (aspect ratio) $\ell =L/D$, gradient $\nabla   \rightarrow  D \nabla $ and chiral strength $ \bar{\varepsilon}_{c}  \rightarrow D^{-3}  \bar{\varepsilon}_{c} $. With the help of the Gaussian averages of the angular terms calculated in \cite{odijkelastic} we obtain the following asymptotic expression for the reference excess free energy:
\begin{align}
 \frac{ F_{ex}}{A}  =  K_{0} & \sim  \rho_{0}^{2} \ell  \left( \frac{\pi}{\sigma} \right )^{\frac{1}{2}}
\end{align}
Combining this with the  orientational free energy, $ F_{\rm id}/A \sim \rho_{0}  (\ln \sigma - 1)$ for freely rotating uniaxial particles and minimising with respect to $\sigma$ yields a common quadratic scaling of the Gaussian variational parameter with density, namely $ \sigma \sim \frac{\pi}{4} \rho_{0}^{2} \ell^{2} $ \cite{Vroege92}. Employing the Gaussian averages listed in the Appendix and applying some algebraic manipulation we can work out the explicit density-dependence of the nematic alignment, torque-field, twist and bend elastic constants, respectively:
\begin{align}
 K_{0}  & \sim  2 \rho_{0} \nonumber \\ 
 K_{t}  & \sim -\rho_{0}^{2}  \bar{\varepsilon}_{c} \nonumber \\ 
 K_{2}  & \sim  \frac{\rho_{0} \ell^{2}}{12}   \nonumber \\ 
 K_{1}   & \sim  \frac{17 \rho_{0} \ell^{2}}{24} = \frac{17}{2} K_{2}   
  \label{kexp}
\end{align}
These results summarise the main findings of this study. We conclude that the moduli of rodlike particle confined to a membrane are different from those of of 3D bulk nematic fluid.  In the limit of asymptotic alignment, the splay-to-twist ratio of a bulk fluid  \cite{odijkelastic}  was predicted to scale as $K_{1}/K_{2} \sim 3 $ whereas a much higher ratio $K_{1}/K_{2} \sim 17/2$ is found for the membrane. The bend-to-twist ratio for a hard rod nematic fluid was found to be proportional to the degree of nematic alignment $K_{3}/K_{2} \sim \sigma \gg 1$ \cite{odijkelastic} where $\sigma$ is steered by the rod concentration.  The curvature-to-twist elasticity of a membrane turns out to be smaller than unity $K_{3}/K_{2} \sim 1/4$ and independent of the rod concentration.  In other words, rods confined to a membrane experience a much stronger resistance to splay fluctuations whereas  bend fluctuations are far less penalised compared to a 3D nematic fluid.  Since the splay modulus is about an order of magnitude larger than the twist elasticity, we expect director deformations whereby rods tilt along the radial vector of the membrane to be of marginal importance.    Fluctuations in the splay angle $\psi$ can therefore be safely  ignored by keeping $\psi =0$ in \eq{tilt}.  

With the scaling expressions for the elastic moduli at hand we proceed toward analyzing the continuum expression for the twist-bend elastic free energy which takes the following form:
\begin{align}
\frac{ F_{mem}}{K_{2}}  &= \int  d \brr \left [  \left ( \partial_{R} \varphi + \frac{\sin 2 \varphi}{2R} + q_{0} \right )^{2}  + \frac{K_{3}}{K_{2}}\frac{ ( \sin \varphi )^{4}}{R^{2}} \right . \nonumber \\ 
& \left .  + \frac{1} {\lambda_{t}^{2}} ( \sin \varphi)^{2} \right   ]
\label{fmem}
\end{align}

\begin{figure}
\begin{center}
\includegraphics[width= 0.6 \columnwidth]{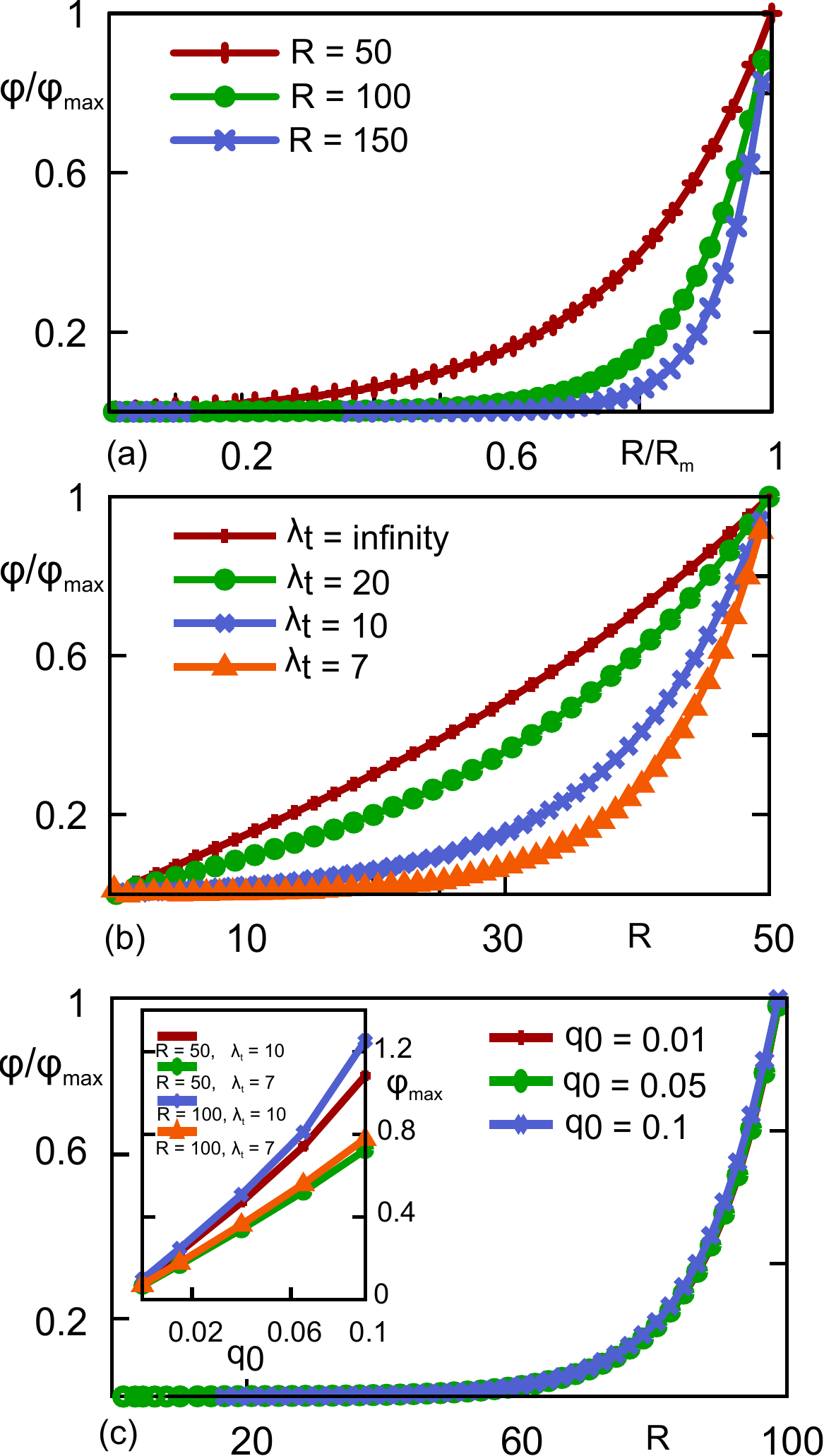}
\caption{ \label{fig2} Typical twist angle profiles plotted versus the membrane distance from the core obtained from minimising \eq{fmem}.  The curves are normalised in terms of the maximum twist angle $\varphi_{\rm max}$ at the membrane edge and correspond to (a) different membrane radii $R_{m}$ at a fixed pitch $q_{0} = 0.05$ and twist penetration depth $\lambda_{t} = 10 $, (b) various $\lambda_{t}$ at fixed $R=50$ and $q_{0}=0.05$, and (c) different values of the chiral pitch $q_{0}$ at fixed $R=100$ and $ \lambda_{t} = 10$. The inset displays the evolution of the edge twist angle as a function of the pitch $q_{0}$ for the parameters indicated.   }
\end{center}
\end{figure}

\noindent This expression is similar to the one discussed in Ref. \cite{barry_jpcb2009} and is defined in terms of the inverse  pitch  $q_{0} = -K_{t}/K_{2} \sim 12 \rho_{0} \bar{\varepsilon}_{c} /\ell^{2}$ quantifying the chiral strength, and bend-to-twist elasticity $K_{3}/K_{2} = 1/4$. We remark that the membrane elastic moduli are strictly 2D quantities with dimension energy. The last term embodies, albeit in a heuristic way,  the effect of depletion attraction due to the non-adsorbing polymer  in terms of a simple free energy $ F_{dep}  \sim a \int d \brr  (\sin \varphi )^{2}  + {\rm cst}$  where $a$ is a tilt energy density (per unit area) related to the osmotic pressure of the polymer reservoir and polymer radius of gyration. A quantitative relation for $a$ could be established using common concepts from free-volume theory \cite{lektuin2011} which we will not pursue here. While more sophisticated expressions for the depletion free energy have been proposed in Ref. \cite{kang_sm2016} the simple sine squared contribution is chosen here for simplicity and is in line with de Gennes' original treatment of twist expulsion towards the edges or around defects of smectic layers in analogy with superconductors \cite{gennes-prost, barry_jpcb2009}. It captures the basic trend that the local director tilting away from the membrane normal compromises the free volume experienced by the non-adsorbing polymer  thereby inducing a free energy penalty. Out-of-plane fluctuations where the rod centre-of-mass departs from the 2D plane turn out negligible for strongly slender rods $\ell \gg 1$   and are not included \cite{kang_sm2016}.
Ignoring the curvature terms 
$R \rightarrow \infty $ and considering a smectic layer on an infinite half-plane enables an analytical minimisation of the free energy in terms of the twist penetration depth $\lambda_{t}  =  \sqrt{K_{2}/a} $ \cite{gennes-prost,barry_jpcb2009}. For the circular membrane, a simple simulated-annealing Monte Carlo algorithm can be employed to minimise the free energy with respect to the twist angle $\varphi(R)$ for any given triplet of length scales, namely the bulk pitch $q_{0}^{-1} $, twist penetration depth $\lambda_{t} $ and membrane radius $R_{m}$. With the twist elastic modulus and chiral amplitude being microscopically defined \eq{kexp} a simple one-parameter fitting procedure can be used to determine the depletion strength $a$ and the twist penetration depth $\lambda_{t}$. 

An overview of the results is shown in \fig{fig2}. The twisting becomes more pronounced toward the membrane edge when the twist penetration depth becomes shorter (\fig{fig2}b) as expected,  but also when membrane size grows larger (\fig{fig2}a).  Increasing the chiral amplitude through the pitch $q_{0}$ merely enhanced the maximum twist angle while keeping the overall shape of the twist angle profile largely unchanged (\fig{fig2}c).
It is fairly straightforward to generalise \eq{fmem} using the full parameterisation \eq{tilt} to account for the splay contributions that are proportional to $\psi$ and its gradients.   Performing a Monte Carlo minimisation of the membrane free energy for the coupled angles $\psi (R)$ and $\varphi(R)$ we find that that splay angle remains negligibly small  across the membrane so that the omission of splay effects seems fully justified a posteriori.

\section{Conclusions}

In this study we have calculated the elastic moduli of a smectic membrane composed of elongated rods in an effort to arrive at more a judicious modelling of the mesoscopic twist profile imparted by chiral rods such as {\em fd} virus particles assembled into a smectic film \cite{dogic-fraden_fil,Gibaud2014}.
While the focus in this study has been on strictly hard rods, assuming the presence of soft chiral forces to be a mere perturbative effect not influencing the elasticity of the membrane,  the theory can be easily extended towards more general rod potentials including attractive contributions, for instance, those mediated by any type of short-range bonding or patchy potential. A simple van der Waals high-temperature expansion, along the lines of the chiral amplitude discussed in Sec. II-B, usually suffices to capture the main impact of particle softness and to gauge the effect of temperature on the elastic properties \cite{gelbart1996}.

 Microscopic knowledge of the elastic and chiral amplitude is the key to devising more appropriate models to predict the structure-property relation of smectic membranes. For example, particle-level approaches similar to the one presented in this work may be explored to model chiral ribbons \cite{Gibaud2014,Srivastava1355} or, more generally, self-assembled polymeric morphologies where the interplay between twist and elasticity plays a central role \cite{Ikkala2407}. The quantitative merits of the scaling expressions derived in this study are currently being tested against computer simulations based on strongly elongated spherocylinders mixed with  penetrable hard spheres acting as a depletion agent with a twisted mesostructure mediated by an additional controllable chiral potential similar to \eq{uchi}. The findings of this study will be published in a future paper.  
\acknowledgements

This work was supported by a grant overseen by the French National Research Agency (ANR) as part of the ``{\em Jeunes Chercheuses et Jeunes Chercheurs (JCJC)}"  Program (ANR-CE32-UPSCALE). The authors are grateful to Anja Kuhnhold and Tanja Schilling (Freiburg University) for helpful discussions and for providing the snapshot in Fig. \ref{fig1}.

\section*{Appendix}

For the calculation of the asymptotic expressions of the elastic moduli we have used the following Gaussian averages  deduced from Ref. \cite{odijkelastic}:
\bea
\langle \langle  |\gamma| \theta_{1}^{3}  \theta_{2} \cos \Delta \phi \rangle \rangle \sim -\frac{7}{4} \pi^{\frac{1}{2}} \sigma^{-\frac{5}{2}} \nonumber \\ 
\langle \langle  |\gamma| \theta_{1}^{2}  \theta_{2}^{2} \cos 2 \Delta \phi \rangle \rangle \sim \frac{1}{4} \pi^{\frac{1}{2}} \sigma^{-\frac{5}{2}} \nonumber \\ 
\langle \langle  |\gamma| \theta_{1}^{2}  \theta_{2}^{2} \rangle \rangle \sim \frac{23}{4} \pi^{\frac{1}{2}} \sigma^{-\frac{5}{2}} 
 \eea

\bibliographystyle{apsrev}
\bibliography{refs}

\end{document}